\documentclass[aip,cp,amsmath,amssymb,reprint]{revtex4-2}
\usepackage{graphicx}
\usepackage[utf8]{inputenc}
\usepackage[T1]{fontenc}
\usepackage{amsthm,amsfonts}
\usepackage{mathrsfs}
\usepackage{mathptmx}
\usepackage{bm}
\usepackage[retainorgcmds]{IEEEtrantools}
\usepackage{siunitx}
\usepackage{hyperref}
\usepackage{commands}
\usepackage{color}
\newcommand{\cor}[1]{\textcolor{red}{#1}}
\begin{document}
% Title portion
\title{Self-consistent Modeling of Inductively Coupled Plasma Discharges}
\author{Alessandro Munaf\`{o}} % Write as First name Surname
 \email[Corresponding author: ]{munafo@illinois.edu}
\author{Sanjeeev Kumar}%
 \email{sanjeeev4@illinois.edu}
\affiliation{\cor{
Center for Hypersonics and Entry Systems Studies (CHESS),
University of Illinois at Urbana-Champaign,
Ceramics Building\char`,{}  105 South Goodwin Avenue\char`, {} Urbana\char`, {} IL 61801 (USA)}}
\author{Marco Panesi}
\email{mpanesi@illinois.edu}
\begin{abstract}
The purpose of this work is the development of a self-consistent multi-physics modeling framework for ICP discharges. Unlike a monolithic approach, the hydrodynamics and electromagnetic field are handled by separate solvers, all developed within the \emph{Center for Hypersonics and Entry Systems Studies} (CHESS) at the University of Illinois. Hydrodynamics is modeled using \textsc{hegel}, a finite volume solver for non-equilibrium plasmas. This solver is interfaced with the \textsc{plato} library, which is responsible for evaluating all plasma-related quantities (\emph{e.g.}, thermodynamic and transport properties). The electric field is handled by \textsc{flux}, a finite element solver. Coupling is realized using the \textsc{preCICE} open-source library. Applications are here presented and discussed to demonstrate the effectiveness of the proposed modeling strategy.
\end{abstract}
\maketitle
\section{INTRODUCTION}\label{sec:intro}
Inductively coupled plasmas (ICPs) have a broad range of applications which include spray processes \cite{fauchais2004understanding}, waste treatment \cite{heberlein2008thermal}, arc welding \cite{murphy2009modelling}, plasma cutting \cite{colombo2009high}, nanopowder fabrication \cite{shigeta2011thermal}, and testing of thermal protection system (TPS) materials for atmospheric entry vehicles.

In the above applications, plasmas are often generated via a suitably designed torch. In its simplest configuration, a plasma torch consists of a quartz tube surrounded  by an  inductor  coil   made  of  a  series  of current-carrying rings. The radio-frequency currents running through the inductor induce toroidal currents in the  gas,  which is heated because of Ohmic dissipation \cite{Reed_1961,boulos1994thermal}. If the energy supplied is large enough, the gas in the torch can undergo ionization and attain temperatures up to, or above, \SI{10000}{\kelvin}. Since the heating occurs via electromagnetic induction, ICPs are essentially contamination-free. This is not the case in arc-jet facilities where material fragments resulting from electrode erosion may alter the plasma composition with undesirable effects on diagnostics techniques (\emph{e.g.}, spectroscopy).  

Inductively coupled plasmas at or near atmospheric pressures are often referred to as thermal plasmas \cite{boulos1994thermal} since the large collision rates among free electrons and heavy particles (\emph{i.e.}, atoms and molecules) ensure that their temperatures are nearly equal. On this basis, large-pressure ICPs are often modeled assuming local thermodynamic equilibrium (LTE). This choice is attractive from the computational point of view as the fluid governing equations remain the global mass, momentum, and energy balance relations \cite{Boulos_1976,Most_1984,Most_1989,Watanabe_1990,bernardi2003three,colombo2011three}.\footnote{This is true as long demixing is neglected.} The only complication, which results from temperature and pressure dependence of thermodynamic and transport properties, may be tackled using suitable look-up tables or curve fits \cite{Rinaldi_JCP_2014}.  

However, there are situations where the LTE assumption breaks down. Aside from low-pressure ICPs, non-local thermodynamic equilibrium (NLTE) effects may be important even around atmospheric pressure. This is the case, for instance, of the fringe region of a plasma jet where the cold chamber gas is entrained by the hot laminar core \cite{pfender1991entrainment}. The ensuing mixing, which eventually leads to transition to turbulence, is inherently a non-equilibrium process involving diffusion along with recombination and de-excitation. Also, even though the non-equilibrium effects may be negligible at the torch exit, the plasma state may still be affected by NLTE during the discharge \cite{Zhang_POP_ICP_2016,Munafo_JAP_2015}. Under these circumstances, LTE simulations tend to overestimate temperatures and lead to significant errors in the chemical composition and size of the plasma volume. These facts must be considered when comparing simulations with experiments.

As it may be deduced from the above discussion, modeling an ICP discharge or wind tunnel is inherently a multi-physics problem that requires coupling between plasma dynamics and electromagnetic phenomena. Moreover, it is worth recalling that an actual ICP facility is always characterized by a certain degree of unsteadiness due to hydrodynamic instabilities, turbulence, and/or arc restrikes \cite{Shigeta_JPhysD_2016}. Despite this, simulations are primarily performed under steady state using engineering turbulence models (\emph{e.g.}, RANS).  

This paper's purpose is to develop a computational framework for ICP discharges. Unlike a monolithic approach, modeling the plasma hydrodynamics and the electromagnetic field relies on separate in-house solvers. Hydrodynamics is modeled using \textsc{hegel} \cite{Munafo_JCP_2020}, a finite volume solver for LTE/NLTE plasmas. Instead, the electric field simulation is accomplished via \textsc{flux}, a finite element solver \cite{Kumar_RGD32} developed on top of the \textsc{mfem} library \cite{mfem,mfem-web}. Coupling is realized using \textsc{preCICE} \cite{preCICE} open-source library.

The paper is structured as follows. First, the physical model is introduced. This is followed by a brief description of the computational framework. Applications to LTE/NLTE plasma discharges are then presented and discussed. Finally, conclusions and future work are outlined.
\section{PHYSICAL MODEL}\label{sec:phys}
\subsection{Plasma}\label{sec:phys_plasma}
The plasmas treated in this work are made of free electrons, neutrals, and ions, all modeled as ideal gases (\emph{e.g.}, no pressure ionization \cite{Zeldovich_book_1967}). The plasma constituents/species are stored in set $\Sset = \{\elec\} \cup \Sseti{h}$, where the symbol $\elec$ denotes free-electrons. The heavy-particle subset $\Sseti{h}$ contains atoms and molecules: $\Sseti{h} = \Sseti{a} \cup \Sseti{m}$. Here the word species may refer to chemical components such as $\mathrm{N}_2$ or $\mathrm{NO}^+$ when considering a multi-temperature (MT) formulation \cite{Park_book,Degrez_JPhysD_2009,Kotov_JCP_2014}, or individual bound-states/groups (\emph{e.g.}, $\mathrm{N}(i)$) for a state-to-state (StS) \cite{Capitelli_book,Bultel_PRE_2002,colonna_2006,Kim_JTHT_2009,Panesi_JTHT_2009,Panesi_JTHT_2011,Pietanza_SpActa_2010,Munafo_EPJD_2012,Munafo_POP_2013,Panesi_2013_BoxRVC,Annaloro_2013,Panesi_PRE_2014,Bender2015,Macdonald_PRF_2016,Luo_JChemPhys_2017,Jo_JPhysChemA_2022} or grouping approach \cite{colonna_JTHT_2006,Munafo_PRE_2014,Munafo_POF_2014,Yen_2015,Munafo_POF_2015,Munafo_ApJ_2017,Sahai_JChemPhys_2017,Mac_JChemPhys_2018I,Mac_JChemPhys_2018II,Sahai_PRF_2019,Campoli_2019,Sharma_PhysRevE_2020,Venturi_JPhysChemA_2020,Venturi_JPhysChemA_2020II,Kim_HMT_2021,Zanardi_2022,Kosareva_2021,Kosareva_2022}. In both situations, heavy-particles and free-electrons are assigned distinct translational temperatures ($\Th$ and $\Te$, respectively) to account for possible thermal non-equilibrium resulting from inefficient energy transfer in electron-heavy collisions.      

The NLTE governing equations, along with constitutive relations for thermodynamics, transport, and kinetics, are obtained via the Chapman-Enskog (CE) expansion for solving the Boltzmann equation  \cite{Ferziger_book,Giov_book,Kustova_book,Devoto_1966,Devoto_1967,Magin_JCP_2004,Magin_PRE_2004,Bruno_POP_2010,Bruno_book}). Here the expansion is stopped at first-order, yielding Newton and Fourier's laws for viscous stresses and heat-fluxes, respectively, in the case of a pure gas made of particles with no internal energy. Kinetic processes (\emph{e.g.}, ionization) are treated assuming a Maxwellian reaction regime \cite{Kustova_book,Giov_book,Bruno_book}.     
\paragraph{Thermodynamics} In view of the ideal gas assumption, the plasma pressure follows from Dalton's law, $p = \ph + \pe$, where the partial pressures of the free-electrons and heavy-particles are, respectively, $\ph = \nbh \kboltz \Th $ and $\pe = \nbe \kboltz \Te$, with $\kboltz$ being Boltzmann's constant. The symbols $\nbe$ and $\nbh$ stand, respectively, for the number density of free-electrons and heavy-particles, with $\nbh = \sum_{s \in \Sseti{h}} \nbi{s}$. Upon introducing the mole fractions $\Xsi{s} = \nbi{s}/n$, the number density may be retrieved from the pressure as:  
\be
n = \fr{p}{\kboltz \Th \left[1 + \Xelec \left(\Te/\Th - 1 \right) \right]}.
\ee
The plasma density is $\rho = \sum_{s \in \Sset} \rhoi{s}$, where the partial densities are $\rhoi{s} = \mi{s} \nbi{s}$, with $\mi{s}$ being the (particle) mass of $s$.

The energy per unit-mass of the individual species reads \cite{Liu_Vinokur_JCP_1989,Grossman_Cinnella_JCP_1990}:
\be
e_s = 
\begin{cases}
\eij{s}{\mathrm{tr}}{\Te},\quad\quad\quad\quad\quad\quad\,\, s = \elec, \\
\\
\eij{s}{\mathrm{tr}}{\Th} + \estari{s} + \eif{s}, \quad\,\,\,\,\, s \in \Sseti{h}, \\
\end{cases}	
\ee
with the translational contribution given by $\eij{s}{\mathrm{tr}}{T} = 3/2 (\kboltz T/\mi{s})$ \cite{Callen_book}. The symbol $\eif{s}$ denotes the absolute formation enthalpy and accounts for both formation and excitation (when using a StS approach). The remaining term, $\estari{s}$, accounts for the energy of the \emph{thermalized} internal degrees of freedom (\emph{e.g.}, rotation, vibration) stored in the set $\Gset$:
\be
\estari{s} = \estari{s}(\Tiset) = \sum_{g \in \Gset} \estari{sg}(\Tiset), \quad s \in \Sseti{h},
\ee
where $\Tiset$ are the internal temperatures (\emph{e.g.}, vibrational, electronic) of the NLTE formulation being considered (\emph{e.g.}, multi-temperature, grouping).  

With the aid of the above definitions, the energy per unit-mass of the plasma, free-electrons, and the \emph{thermalized} internal degrees of freedom may be written as:
\begin{IEEEeqnarray}{rCL}
\IEEEyesnumber\label{eq:rhoe_nlte}\IEEEyessubnumber*	
e           & = & \sum_{s \in \Sseti{h}} \yi{s} \left[\eij{s}{\mathrm{tr}}{\Th} + \eif{s}\right] + \sum_{g \in \Gset} \tilde{e}_g + \tilde{e}_{\elid}, \\
\eti{g}     & = & \sum_{s \in \Sseti{h}} \yi{s} e^{\star}_{sg}(\Tiset), \quad g \in \Gset, \\
\eti{\elid} & = & \yi{\elid} \eij{\elid}{\mathrm{tr}}{\Te},
\end{IEEEeqnarray}
where the mass fractions are $\yi{s} = \rhoi{s}/\rho$.
\paragraph{Transport} The application of the CE method yields explicit expressions for transport fluxes. In the first-order approximation, these fluxes are linearly related to gradients of macroscopic quantities such as velocity and temperatures, the proportionality factors being the transport properties. The former are given by \emph{bracket} integrals which are practically evaluated via a Sonine-Laguerre polynomial expansion \cite{Ferziger_book,Giov_book}.  Here secondary effects such as thermal diffusion are neglected. These are, however, available in the ICP modeling framework. Only the main results are quoted in the following (details may be found in the above references on Kinetic Theory). 

Viscous stresses are given by Newton's law:
\be
\mathsf{\tau} = \mu \left[\nabla \mathbf{v} + \left(\nabla \mathbf{v}\right)^{\trans} - \fr{2}{3} \left( \nabla \cdot \mathbf{v} \right) \mathsf{I} \right],
\ee
where the $\trans$ superscript denotes the transpose, whereas $\mathsf{I}$ stands for the identity tensor. In the first Sonine-Laguerre approximation, the dynamic viscosity reads \cite{Giov_book}:
\be \label{eq:mu_nlte}
\mu = \mathbf{z}^{\mu}_{\mathrm{h}} \cdot \mathbf{X}_{\mathrm{h}}, 
\ee  
where the vector $\mathbf{X}_{\mathrm{h}}$ stores the mole fractions of heavy particles. The entries of $\mathbf{z}^{\mu}$ are solutions of the linear algebraic system:
\be
\mathbf{G}^{\mu} \mathbf{z}^{\mu}_{\mathrm{h}} = \mathbf{X}_{\mathrm{h}},
\ee
with $\mathbf{G}^{\mu}_{\mathrm{h}}$ being the heavy-particle subsystem symmetric transport matrix for viscosity. Free-electrons do not contribute to viscous stresses due to their small mass \cite{Devoto_1966}. 

The diffusion velocities $\mathbf{U}_s$ satisfy Stefan-Maxwell's equations:
\begin{IEEEeqnarray}{rCL}
\IEEEyesnumber\label{eq:sm_nlte}\IEEEyessubnumber*
\sum_{p \in \Sset} \Gdifij{p}{\elid} \mathbf{U}_p - \kdif{\elid} \fr{\Th}{\Te} \mathbf{E} & = & -  \mathbf{d}^{\prime}_{\mathrm{e}} \fr{\Th}{\Te}, \\
\sum_{p \in \Sset} \Gdifij{p}{s} \mathbf{U}_p - \kdif{s} \mathbf{E} & = & -\mathbf{d}^{\prime}_s, \quad s \in \Sseti{h},
\end{IEEEeqnarray}
where the $\Gdifij{s}{p}$ are the entries of the symmetric Stefan-Maxwell matrix $\mathbf{G}^U$, whereas $\mathbf{E}$ is the electric field. The latter accounts, in general, for both external sources and charge distribution within the plasma (\emph{i.e.}, self-induced electric field). The $\kdif{s}$ are defined as $\smash{\kdif{s} = (\Xsi{s} \Qi{s} - \yi{s}Q)/\kboltz \Th}$, where the plasma charge is $\smash{Q = \sum_{s \in \Sset} \Xsi{s}\Qi{s}}$, where $\Qi{s}$ denotes the charge of $s$. The modified diffusion driving forces in Eqs. \eqnref{eq:sm_nlte} are:  
\be \label{eq:mod_driv_nlte}
\mathbf{d}^{\prime}_s = \left(\fr{p}{n \kboltz \Th}\right) \nabla \Xsi{s} + \left(\fr{\Xsi{s} - \yi{s}}{n \kboltz \Th}\right) \nabla p , \quad s \in \Sset.
\ee
It is worth mentioning that the $\mathbf{d}^{\prime}_s$ are not independent since $\sum_{s \in \Sset} \mathbf{d}^{\prime}_s = \mathbf{0}$, as shown by a direct calculation. For the ICP simulations considered in this work, Eq. \eqnref{eq:mod_driv_nlte} may be simplified as follows. Since the pressure is essentially constant, the second term on the right-hand side may be dropped. Additionally, if the difference between heavy-particle and free-electron temperatures is not too large, the pressure becomes $p \simeq n \, \kboltz \Th$, which leads to $\mathbf{d}^{\prime}_s \simeq \nabla \Xsi{s}$.  
 
The diffusion velocities are found by solving Eqs. \eqnref{eq:sm_nlte} along with mass conservation and ambipolar diffusion constraints which, when combined together, give $\sum_{s \in \Sset} \kdif{s} \mathbf{U}_s = \mathbf{0}$ \cite{Magin_JCP_2004}. The solution of Eqs. \eqnref{eq:sm_nlte}, along with the previous relation, yields both diffusion velocities and (ambipolar) electric field.  

For a multi-component NLTE plasma, the total, internal, and free-electron heat-flux components account for both heat conduction and mass diffusion and read:
\begin{IEEEeqnarray}{rCL}
\IEEEyesnumber\label{eq:heat_nlte}\IEEEyessubnumber*
\mathbf{q} & = &- \lamh \nabla \Th + \sum_{s \in \Sseti{h}} \mathbf{J}_s \left[\hij{s}{\mathrm{tr}}{\Th} + \eif{s}\right] +  \sum_{g \in \Gset} \mathbf{\tilde{q}}_g + \mathbf{q}_{\mathrm{e}},  \\
\mathbf{\tilde{q}}_g & = & - \tilde{\lambda}_g\nabla \Tim{g} + \sum_{s \in \Sseti{h}} \mathbf{J}_s \, \estari{sg}(\Tiset), \quad g \in \Gset \\
\mathbf{q}_{\mathrm{e}} & = & - \lame \nabla \Te + \mathbf{J}_{\mathbf{e}} \, \hij{\elid}{\mathrm{tr}}{\Te},
\end{IEEEeqnarray}
where the translational enthalpies and the mass diffusion fluxes are $\hij{s}{\mathrm{tr}}{T} = \eij{s}{\mathrm{tr}}{T} + \kboltz T/\mi{s}$ and $\mathbf{J}_s = \rhoi{s} \mathbf{U}_s$, respectively.

The translational conductivity of heavy particles is evaluated in the second Sonine-Laguerre approximation:
\be \label{eq:mu_lamh_nlte}
\lamh = \mathbf{z}^{\lambda}_{\mathrm{h}} \cdot \mathbf{X}_{\mathrm{h}},
\ee
where, in analogy with viscosity, the entries of $\mathbf{z}^{\lambda}_{\mathrm{h}}$ follow from the solution of the linear system:
\be
\mathbf{G}^{\lambda}_{\mathrm{h}} \mathbf{z}^{\lambda}_{\mathrm{h}}  = \mathbf{X}_{\mathrm{h}},
\ee
with $\mathbf{G}^{\lambda}_{\mathrm{h}}$ being the heavy-particle subsystem symmetric transport matrix for thermal conductivity \cite{Giov_book}. For free-electrons, a third-order Sonine approximation is instead considered \cite{Devoto_1966,Magin_JCP_2004}:
\be
\lame = \fr{75}{64} \kboltz \Xelecs \sqrt{\fr{2\pi\kboltz\Te}{\melec}} \fr{\Lameeij{2}{2}}{\Lameeij{1}{1}\Lameeij{2}{2} - \Lameeij{1}{2}\Lameeij{2}{1}}, 
\ee 
where the $\Lameeij{i}{j}$ are the entries of the electron subsystem symmetric transport matrix \cite{Munafo_JCP_2020,Magin_JCP_2004}. The contribution to thermal conductivity of the internal \emph{thermalized} degrees of freedom, $\tilde{\lambda}_g$, is modeled based on the generalized Eucken correction \cite{Giov_book,Bruno_book}. 
 
The conduction current within the plasma is modeled based on Ohm's law, $\mathbf{j} = \sige \, \mathbf{E}$, where the electrical conductivity is (second Sonine-Laguerre approximation) \cite{Magin_JCP_2004}:
\be\label{eq:sigma_plasma}
\sige = \fr{3}{8} \fr{\Xelecs \Qelecs}{\kboltz \Te} \sqrt{\fr{2\pi\kboltz\Te}{\melec}} \fr{\Lameeij{1}{1}}{\Lameeij{0}{0}\Lameeij{1}{1} - \Lameeij{0}{1}\Lameeij{1}{0}}.
\ee
\paragraph{Kinetics} The NLTE kinetics mechanism adopted in this work includes:
\begin{itemize}
  \item dissociation by heavy-particle and electron impact,
  \item particle and charge exchange (\emph{e.g.}, Zel'dovich reactions),
  \item ionization and excitation by electron impact,
  \item associative ionization and dissociative recombination,
  \item elastic energy transfer in electron-heavy collisions.
\end{itemize}
Radiative processes such as line emission and absorption are not taken into account. 

The mass and energy production terms due to the above processes follow, again, from the CE method. Since a Maxwellian reaction regime is assumed, the source terms are obtained via moments of the collision operator in the Boltzmann equation with the distribution function taken as Maxwell-Boltzmann at the appropriate temperature \cite{Giov_book,Kustova_book}. 
\paragraph{Governing equations} The equations governing the hydrodynamics of the plasmas treated in this work are: 
\be \label{eq:nlte_gov_eq}
\fr{\pa \Cons}{\pa t} + \nabla \cdot \left(\mathsf{F} - \mathsf{D}\right) = \St,
\ee
where $t$ denotes time. The conservative variable and source term vectors, and the inviscid and diffusive flux tensors are: 
\be\label{eq:nlte_gov_eq_vec}
\Cons = \left(
\begin{array}{c}
\rhoi{s} \\
\rho \mathbf{v} \\
\rho E \\
\rho \eti{g} \\
\rho \eti{\elid}
\end{array}
\right),
\quad 
\mathbf{S} = \left(
\begin{array}{c}
\omega_s \\
\left<\mathbf{f}^{\mathrm{L}}\right> \\
\left<\Omega^{\mathrm{J}}\right> \\
\tilde{\Omega}_g \\ 
\left<\Omega^{\mathrm{J}}\right> + \Omega_{\elid} - \pe \nabla \cdot \mathbf{u}
\end{array}
\right), \quad 
\mathsf{F} = \left( 
\begin{array}{c}
\rho \mathbf{v} \mathsf{I} \\
\rho \mathbf{v} \mathbf{v} + p \mathsf{I} \\
\rho \mathbf{v} H \mathsf{I} \\
\rho \mathbf{v} \eti{g} \mathsf{I} \\
\rho \mathbf{v} \eti{\elid} \mathsf{I}
\end{array}
\right) \quad \text{and} \quad
\mathsf{D} = \left(
\begin{array}{c}
-\mathbf{J}_s \mathsf{I} \\
\mathsf{\tau} \\
\left(\mathsf{\tau} \mathbf{v} - \mathbf{q}\right) \mathsf{I} \\
-\mathbf{\tilde{q}}_g \mathsf{I} \\
-\mathbf{q}_{\mathrm{e}} \mathsf{I}
\end{array}
\right),
\ee
for $s \in \mathcal{S}$ and $g \in \mathcal{G}$. The total energy and enthalpy per unit-mass are defined as $E = e + \mathbf{v} \cdot \mathbf{v}/2$ and $H = E + p/\rho$, respectively, with $\mathbf{u} = (u, \, v, \, w)$ being the mass-averaged velocity. The time-averaged Lorentz force, $\smash{\left<\mathbf{f}^{\mathrm{L}}\right>}$, and Joule heating, $\smash{\left<\Omega^{\mathrm{J}}\right>}$, account for the interaction between the plasma and the electromagnetic field (see below). Finally, the mass production terms, $\omi{s}$, and the energy transfer terms, $\tilde{\Omega}_g$ and ${\Omega}_{\elid}$, represent the effects of kinetic processes on the mass and energy balance of the plasma.
\subsection{Electromagnetic field}\label{sec:phys_efield} 
Electromagnetic phenomena are governed by Maxwell's equations. To make the problem tractable, the following assumptions are introduced \cite{Most_1989,David_thesis,VandenAbeele_2000}:
\begin{itemize}
    \item Low-frequency approximation. The inductor frequency, \cor{$f$}, is much smaller than that of the plasma, allowing to rule out both electrostatic and electromagnetic waves.
    \item The plasma is quasi-neutral, unmagnetized, and collision-dominated.
    \item Low magnetic Reynolds number.
    \item Harmonic time-dependence of all electromagnetic quantities: 
        \be\label{eq:time_dep_harm}
           \mathbf{E}(\mathbf{r}, t) = \mathbf{E}_{\mathrm{c}}(\mathbf{r})\exp(\imath \, \omega t),  
          \ee
where the angular frequency is $\omega = 2\pi f$, whereas $\imath$ stands for the imaginary unit. In the above relation, the c subscript denotes a complex quantity (\emph{i.e.}, phasor).
\end{itemize}

The use of the above assumptions in Maxwell's equations leads to the induction equation for the \cor{the electric field phasor}:
\be\label{eq:efield_ampl}
\nabla \times \nabla \times \mathbf{E}_{\mathrm{c}} + \imath \mu_0 \sigma \omega \mathbf{E}_{\mathrm{c}} = -\imath \mu_0 \omega \, \mathbf{j}_s,
\ee
where $\mu_0$ is the vacuum permeability. The electrical conductivity, $\sigma$, is the one of the \cor{plasma} inside the torch, whereas it is assumed zero anywhere else. The  $\mathbf{j}_s$ vector on the right-hand side of Eq. \eqnref{eq:efield_ampl} is the current density contribution from external sources (\emph{i.e.}, inductor coils). 

\cor{Once the electric field is known, the Joule heating and the Lorentz force follow from $\Omega^{\mathrm{J}} = \mathbf{j} \cdot \mathbf{E}$ and $\mathbf{f}^{\, \mathrm{L}} = \mathbf{j} \times \mathbf{B}$, respectively, where the magnetic induction, $\mathbf{B}$, may be retrieved from the electric field via Faraday's law.} Since ICPs operate at frequencies of the order of \si{\mega\hertz}, it is reasonable to assume that over the inductor period  the plasma is effectively subjected to a time-averaged electromagnetic force and energy deposition \cite{David_thesis,VandenAbeele_2000}:
\be \label{eq:Lorentz_Joule_terms}
\left<\mathbf{f}^{\, \mathrm{L}} \right>= \fr{1}{2}\left(\frac{\sigma}{\omega}\right)\cor{\Re{\left[\mathbf{E}_{\mathrm{c}} \times\left(i\nabla \times \mathbf{E}_{\mathrm{c}}\right)^*\right]}} \quad \text{and} \quad
\left<\Omega^{\mathrm{J}}\right> =  \fr{1}{2} \sigma \, \mathbf{E}_{\mathrm{c}} \cdot \mathbf{E}^*_{\mathrm{c}},
\ee
where the $*$ superscript \cor{denotes} the complex conjugate\cor{, whereas $\Re{\left(z\right)}$ stands for the real part of $z$.}

Following Boulos \cite{Boulos_1976}, during the course of a simulation the intensity of the current running through the inductor is updated to match a target value of the power dissipated by Joule heating:
\be
P = \!\! \int \! \left< \Omega^{\mathrm{J}} \right> dv.
\ee
\section{COMPUTATIONAL FRAMEWORK}\label{sec:num}
\cor{As shown in the previous Section, the body force and energy deposition experienced by the plasma depend on the electric field amplitude and gradients which, in turn, are affected by the electrical conductivity of the plasma. To achieve self-consistency, the two set of equations (\emph{i.e.}, plasma and electromagnetic field) must be therefore solved together. Here this is accomplished by coupling two separate solvers. Compared to a monolithic approach, this strategy has the advantage of reducing software complexity and maintenance work. Moreover, one may adopt the most suitable numerical method and algorithm for each sub-problem. The main features of the solvers and the coupling are summarized below.}
\begin{figure}[!htb]
  \centering
  \includegraphics[clip,bb=0 -1 1100 533,keepaspectratio,width=1.0\textwidth]{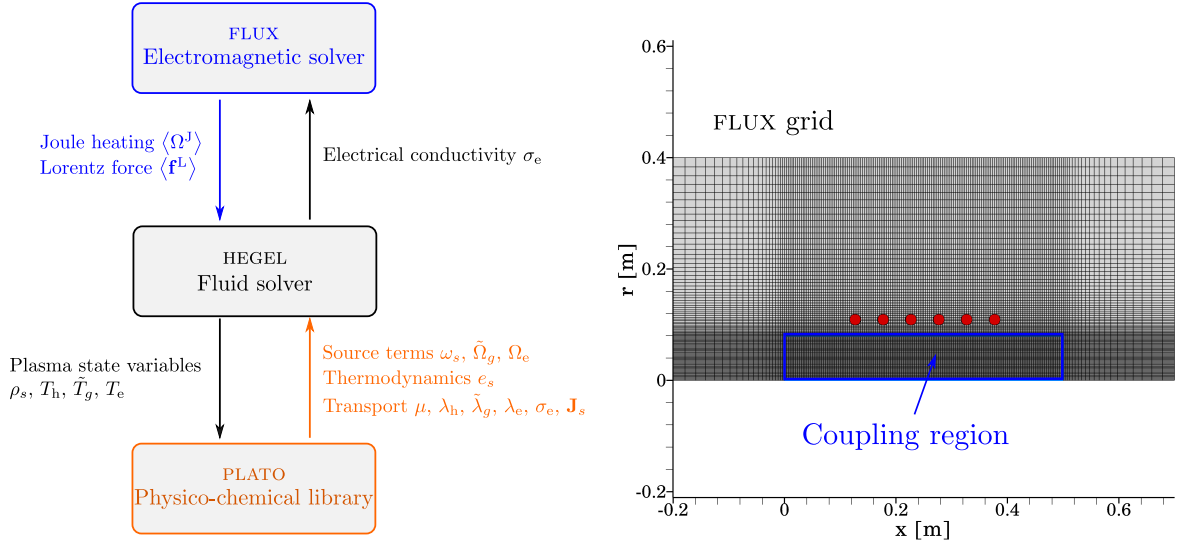}
\caption{Schematic illustrating the coupling between \textsc{hegel} and \textsc{flux} for simulating  LTE and NLTE ICP discharges. The block diagram on the left illustrates the quantities being exchanged. The grid system on the right (coarse grids are shown for the sake of clarity) highlights the region where coupling occurs (\emph{e.g.}, torch). The red circles denote the location of the coils.}\label{fig:coupling}
\end{figure}
\vspace{-1.0cm}
\subsection{Solvers}
\paragraph{Plasma}  
The fluid/plasma solver is \textsc{hegel} (High-fidElity tool for maGnEto-gasdynamics simuLations), a parallel multi-block structured code for LTE and NLTE plasmas written in object-oriented \texttt{Fortran} 2008 \cite{Munafo_JCP_2020,Andrea_JPhysD_2019}. Distribution of data among processes is performed using MPI along with data structures provided by the \textsc{petsc} library \cite{petsc-web-page,petsc-user-ref,petsc-efficient}. The evaluation of thermodynamic and transport properties and source terms is accomplished via the \textsc{plato} (PLAsmas in Thermodynamic nOn-equilibrium) library \cite{Munafo_JCP_2020}.

The governing equations \eqnref{eq:nlte_gov_eq_vec} are discretized in space based on the cell-centered finite volume method. Inviscid fluxes are evaluated using flux functions such as Roe's approximate Riemann solver \cite{Roe_JCP_1981} or the AUSM-family method \cite{AUSMp} along with reconstruction procedures such as MUSCL \cite{muscl}  or WENO \cite{Jiang_Shu_JCP_1996} to achieve high-order accuracy. Diffusive fluxes are computed using Green-Gauss' theorem to determine face-averaged gradients. The space-discretized system of equations is integrated in time via explicit, implicit, or implicit-explicit (IMEX) methods \cite{Munafo_JCP_2020}. 
\paragraph{Electromagnetic field} The equation governing the \cor{electric field phasor} \eqnref{eq:efield_ampl} is solved using \textsc{flux} \cite{Kumar_RGD32}, a \texttt{C++} \textsc{mfem}-based \cite{mfem,mfem-web} mixed finite element solver for time- and frequency-domain electromagnetics. Details are given in a companion manuscript \cite{Kumar_RGD32}.
\subsection{Coupling}
The coupling between \textsc{hegel} and \textsc{flux} is practically realized by means of the \textsc{preCICE} open-source library \cite{preCICE}. \cor{In this work} \textsc{flux} receives from \textsc{hegel} the plasma electrical conductivity, which is then used to compute the electric field. Once this step is completed, the time-averaged Lorentz force and Joule heating are evaluated and sent to \textsc{hegel} (see Fig. \ref{fig:coupling}).
\section{RESULTS}\label{sec:res}
This Section illustrates applications of the developed ICP modeling framework. Here the main purpose is the verification through comparison with data available in the literature. Applications to more complex \cor{scenarios} (\emph{e.g.}, StS modeling) are discussed in a companion paper \cite{Kumar_RGD32}.  
\subsection{LTE ICP torch}
\begin{figure}[!htb]
  \centering
  \includegraphics[clip,bb=0 0 662 168,keepaspectratio,width=0.8\textwidth]{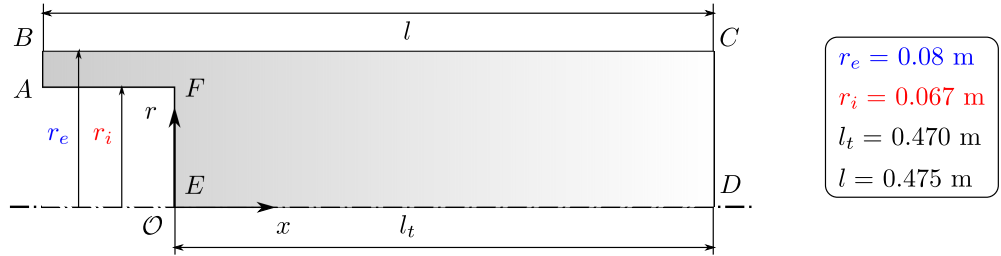}
  \caption{ICP torch with annular injection: geometry and dimensions.}\label{fig:bc}
\end{figure}

The first benchmark consists in computing the two-dimensional axisymmetric flow without swirl in an ICP torch with annular injection (see Fig. \ref{fig:bc}). The working fluid is air, and LTE is assumed. Geometry, operating conditions, and reference solution are taken from Ref. \cite{David_thesis}.   

The boundary conditions are as follows \cite{Zhang_POP_ICP_2016}.
\begin{itemize}
   \item Inlet (AB):
      \be 
		\rho u = \fr{\dot{m}}{\pi \left(r^2_e - r^2_i\right)}, \quad \fr{\pa p}{\pa x} = 0 \quad \text{and} \quad T = T_{\mathrm{in}},
      \ee 	
      where $\dot{m}$ and $T_{\mathrm{in}}$ are, respectively, the mass flow and the temperature of the cold gas being injected. The symbols $r_i$ and $r_e$ stand. respectively, for the inner and outer radii of the injector. 
     \item Centerline (DE):
      \be 
		\fr{\pa \rho}{\pa r} = \fr{\pa u}{\pa r} = \fr{\pa p}{\pa r} = 0 \quad \text{and} \quad v = 0,
      \ee		
   \item Walls (AF and BC):
      \be 
         u = v = 0 \quad \text{and} \quad T = T_{\mathrm{w}},
      \ee		
      where $T_{\mathrm{w}}$ denotes the wall temperature.
   \item Wall (EF)
      \be 
         u = v = 0 \quad \text{and} \quad \fr{\pa T}{\pa x} = 0,
      \ee  
   \item Outlet (CD):
      \be 
         p = p_{\mathrm{a}},
      \ee		
      where $p_{\mathrm{a}}$ is the ambient pressure.
\end{itemize}
The mass flow and ambient pressure are set to \SI{6}{\gram/\second} and \SI{5000}{\pascal}, respectively, whereas the wall and inlet temperatures are both equal to \SI{350}{\kelvin}. The target dissipated power and the frequency of the current running through the inductor are \SI{50}{\kilo\watt} and \SI{0.45}{\mega\hertz}, respectively.

Since the modeling of the plasma formation is out of the scope of this work, the calculation is started by imposing a high-temperature plasma blob in the torch \cite{VandenAbeele_2000}. The fluid governing equations are then marched in time using the backward Euler method \cite{Hirsch_book_volI} along with local time-stepping to accelerate convergence to steady-state \cite{Blazek_book_2015}. To this purpose, the Courant-Friedrichs-Levy (CFL) number is also increased during the course of the simulation. The data exchange between \textsc{hegel} and \textsc{flux} is performed at the end of each fluid time-step via explicit coupling. In the present case, data could also be exchanged every ten fluid \cor{time-steps} without deteriorating the convergence of the numerical solution. 
\begin{figure}[!htb]
  \centering
  \begin{tabular}[b]{c}
    \includegraphics[clip,bb=0 20 592 505,keepaspectratio,width=0.50\textwidth]{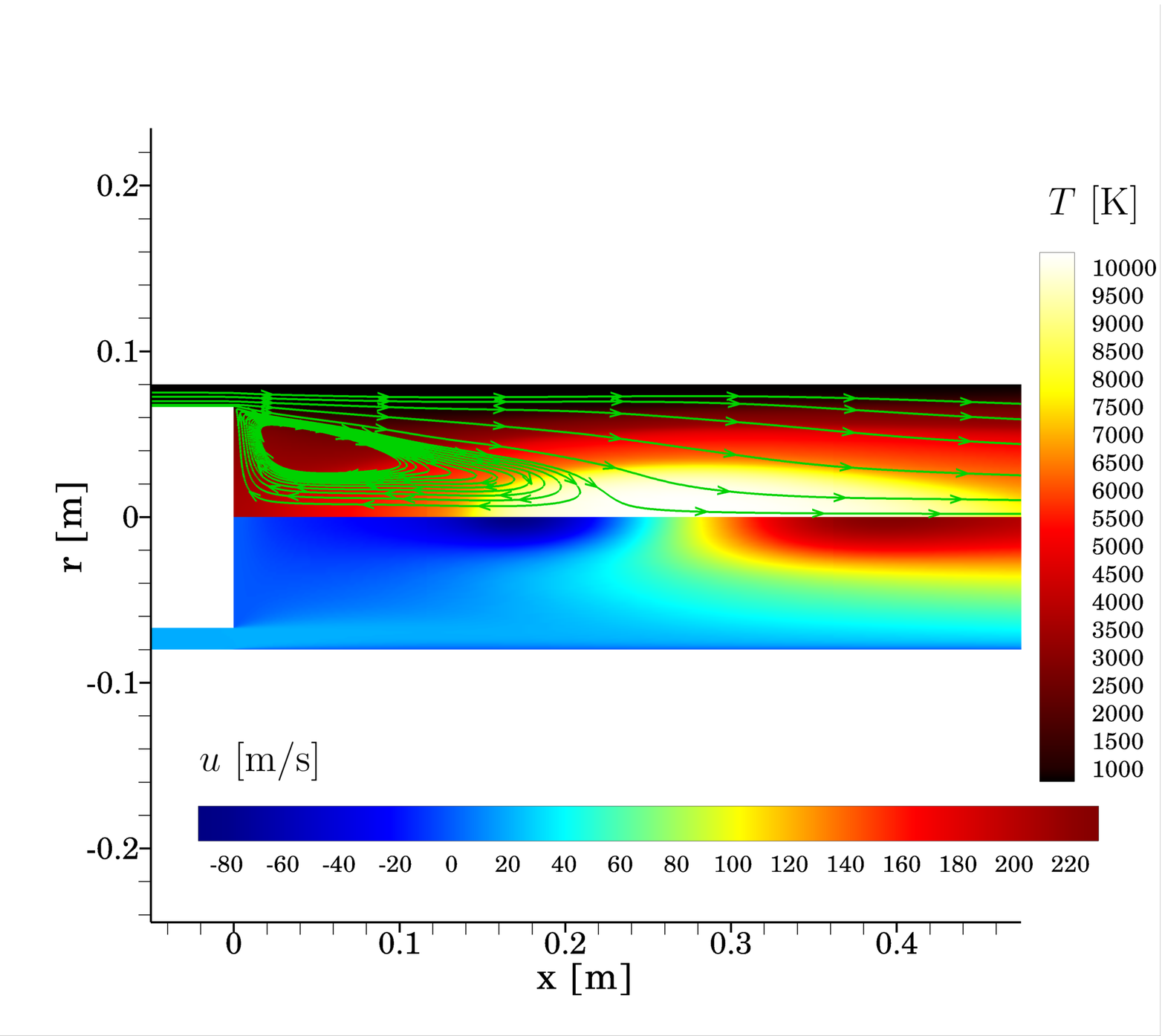} \\
    \small (a) Flowfield.
  \end{tabular} \qquad
  \begin{tabular}[b]{c}
    \includegraphics[clip,bb=0 3 273 192,keepaspectratio,width=0.44\textwidth]{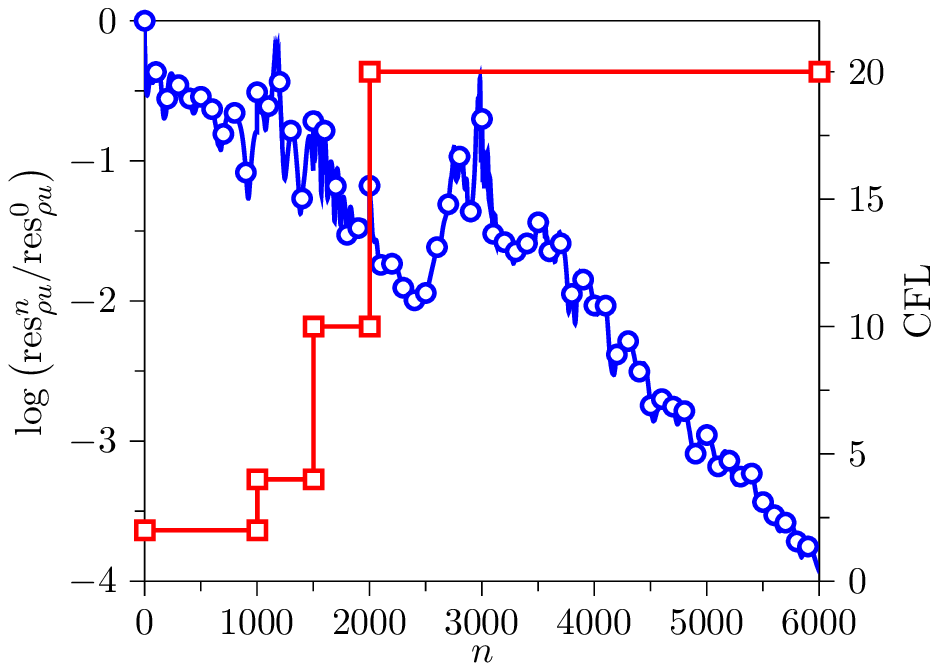} \\
    \small (b) Convergence history.
  \end{tabular}
     \caption{}
	\caption{ICP torch simulation (LTE air plasma): in (a) temperature and axial velocity distributions with streamlines, in (b) line with circles logarithm of axial momentum density residual [see Eq. \eqnref{eq:res_rhou}] normalized with respect to the first iteration, line with squares CFL number ($\dot{m} = \SI{6}{\gram/\second}$, $f = \SI{0.45}{\mega\hertz}$, $P = \SI{50}{\kilo\watt}$, $p_{\mathrm{a}}= \SI{5000}{\pascal}$, $T_{\mathrm{w}} = \SI{350}{\kelvin}$; no swirl).}\label{fig:torch_lte}
\end{figure}
\begin{figure}[!htb]
\includegraphics[clip,bb=0 3 255 195,keepaspectratio,height=0.385\textwidth]{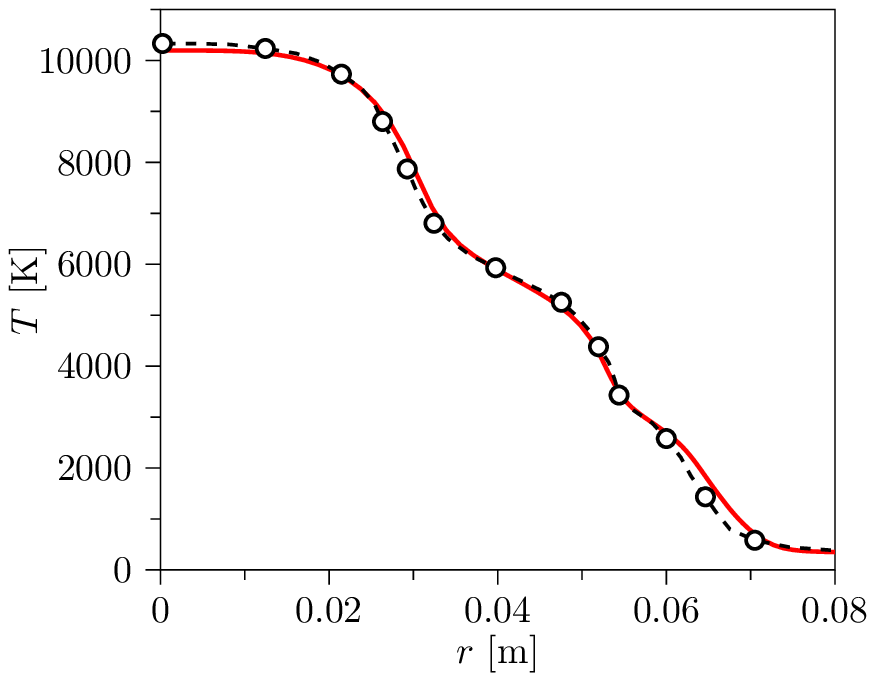} 
\caption{ICP simulation (LTE air plasma): temperature radial profile at $x=\SI{0.265}{\meter}$. Line this work, dashed line with circles solution from Refs. \cite{David_thesis} ($\dot{m} = \SI{6}{\gram/\second}$, $f = \SI{0.45}{\mega\hertz}$, $P = \SI{50}{\kilo\watt}$, $p_{\mathrm{a}}= \SI{5000}{\pascal}$, $T_{\mathrm{w}} = \SI{350}{\kelvin}$; no swirl).}\label{fig:torch_lte_ver}
\end{figure}

Figure \ref{fig:torch_lte} shows the computed axial velocity and temperature distributions, along with the convergence history monitored by plotting the residual of the axial momentum density:
\be\label{eq:res_rhou}
\mathrm{res}^n_{\rho u} = \sqrt{\frac{1}{\mathrm{N_I} \mathrm{N_J}} \sum_{i}^{\mathrm{N_I}} \sum_{j}^{\mathrm{N_J}} \left(\delta \rho u^n_{i, \, j}\right)^2},
\ee	
where the solution increment between time-level $n$ and $n+1$ for cell $(i, \, j)$ is $\delta \rho u^n_{i, \, j} = \rho u^{n+1}_{i, \, j} - \rho u^n_{i, \, j}$, with $\mathrm{N_I}$ and $\mathrm{N_J}$ being, respectively, the number of cells along the axial and radial directions. 

The streamlines on top of the temperature field show the characteristic recirculation eddy resulting from electromagnetic pumping \cite{Most_1989,Watanabe_1990}. The temperature is maximum on the axis, with peak values around \SI{10000}{\kelvin}, as also shown in Fig. \ref{fig:torch_lte_ver} which compares the present results with the reference solution \cite{David_thesis}. Overall, the agreement is very good with minor differences probably due to the use of a different physico-chemical database (\emph{e.g.}, transport collision integrals) and numerical method. The radial temperature distribution is flat close to the axis due to neglecting radiation losses \cite{Munafo_JAP_2015} and undergoes a series of inflection points. These are consequences of local maxima of the total LTE thermal conductivity of air. 
\begin{figure}[!htb]
\includegraphics[clip,bb=10 10 590 500,keepaspectratio,width=0.6\textwidth]{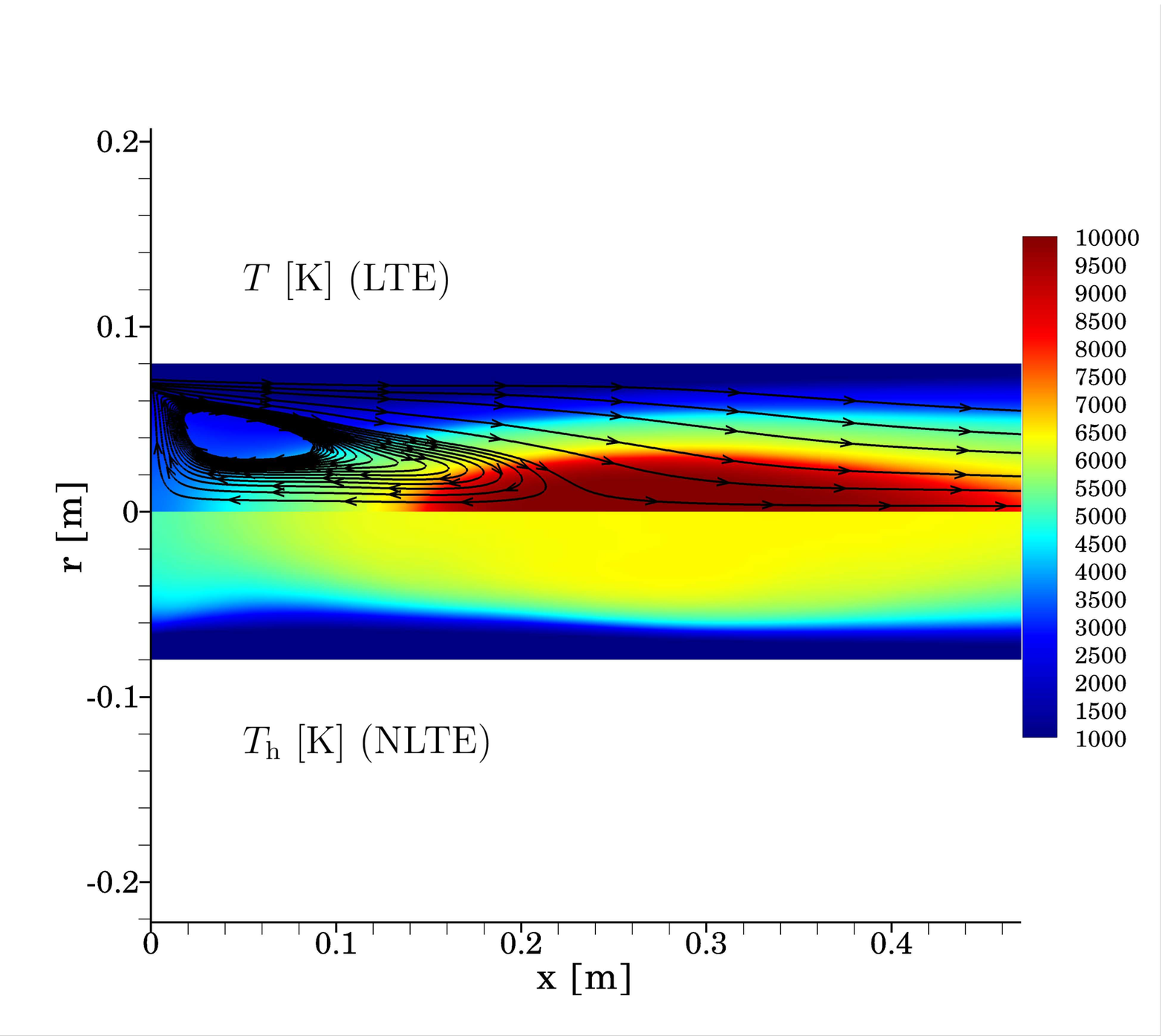}	
\caption{ICP simulation (LTE and NLTE air plasma): top LTE temperature distribution with streamlines, bottom NLTE heavy-particle temperature distribution ($\dot{m} = \SI{6}{\gram/\second}$, $f = \SI{0.45}{\mega\hertz}$, $P = \SI{50}{\kilo\watt}$, $p_{\infty}= \SI{5000}{\pascal}$, $T_{\mathrm{w}} = \SI{350}{\kelvin}$; no swirl).}\label{fig:torch_nlte}
\end{figure}
\subsection{NLTE ICP torch}
After assessing the correct implementation of the LTE formulation of both solvers, the previous LTE simulation was repeated under NLTE conditions. 

The air plasma is made of $\mathrm{N}_2$ and $\mathrm{O}_2$, and their main dissociation and ionization products:
\be
\mathcal{S} = \left\{\mathrm{e}^-,\,\mathrm{N}_2,\,\mathrm{O}_2,\,\mathrm{NO},\,\mathrm{N},\,\mathrm{O},{\mathrm{N}_2}^{+},\,{\mathrm{O}_2}^{+},\,\mathrm{NO}^+,\,\mathrm{N}^+,\,\mathrm{O}^+\right\}.
\ee
Non-equilibrium effects are taken into account based on the Park two-temperature model \cite{Park_1993_Earth} along with Dunn and Kang reaction kinetics scheme \cite{Gnoffo_nasatp_1989}. For the sake of consistency in the verification procedure, the rate controlling temperatures for the various chemical reactions (\emph{e.g.}, dissociation, exchange) are taken from Ref. \cite{David_thesis}. It is important to mention that the NLTE model is built upon using the same database (\emph{e.g.}, thermodynamics, transport) used for the LTE simulation. This ensures the self-consistency of the LTE vs NLTE comparison.    
 
\begin{figure}[!htb]
  \centering
  \begin{tabular}[b]{c}
    \includegraphics[clip,bb=10 10 590 580,keepaspectratio,width=0.47\textwidth]{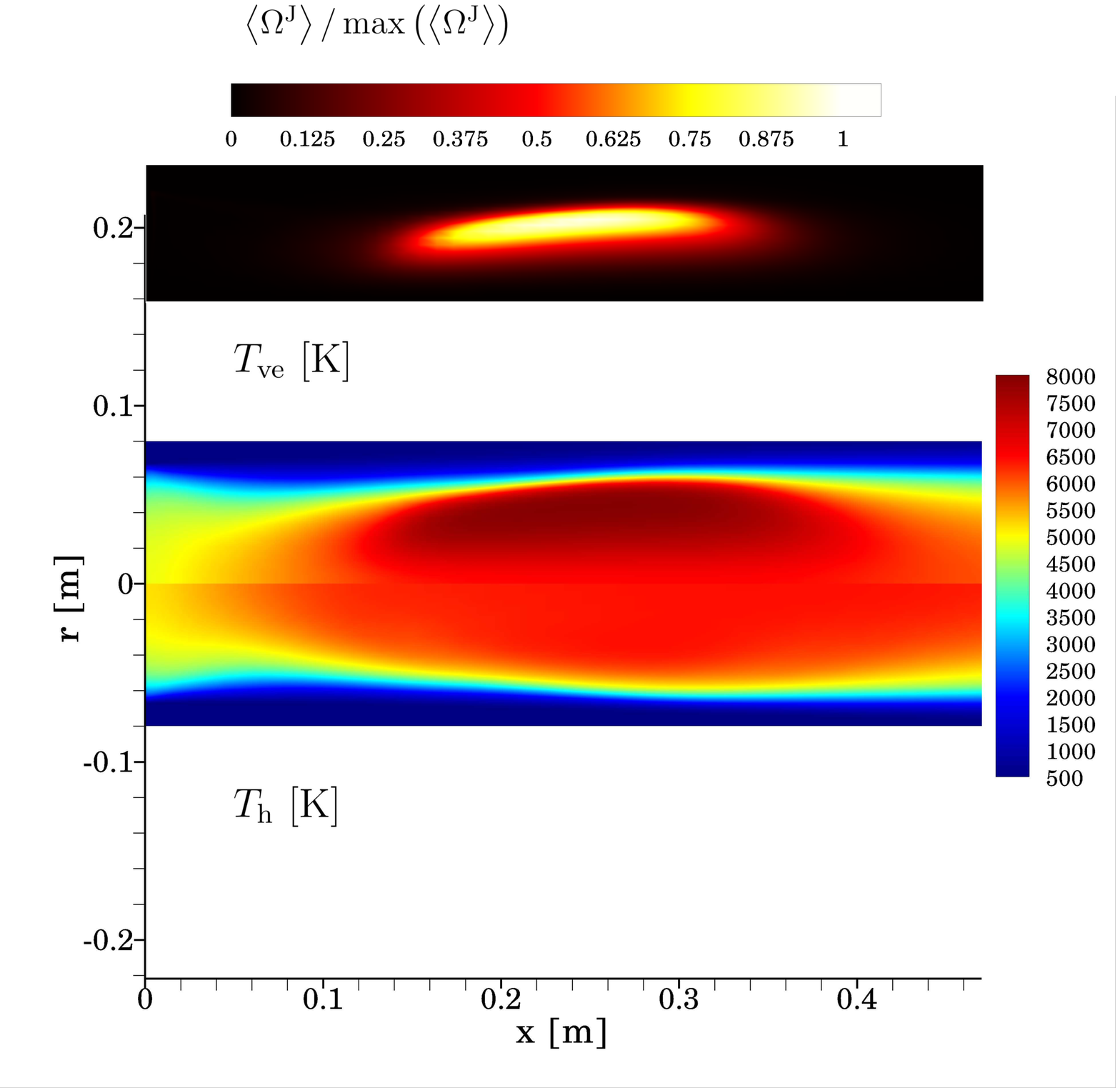} \\
    \small (a) Temperatures and Joule heating.
  \end{tabular} \qquad
  \begin{tabular}[b]{c}
    \includegraphics[clip,bb=0 3 261 195,keepaspectratio,width=0.47\textwidth]{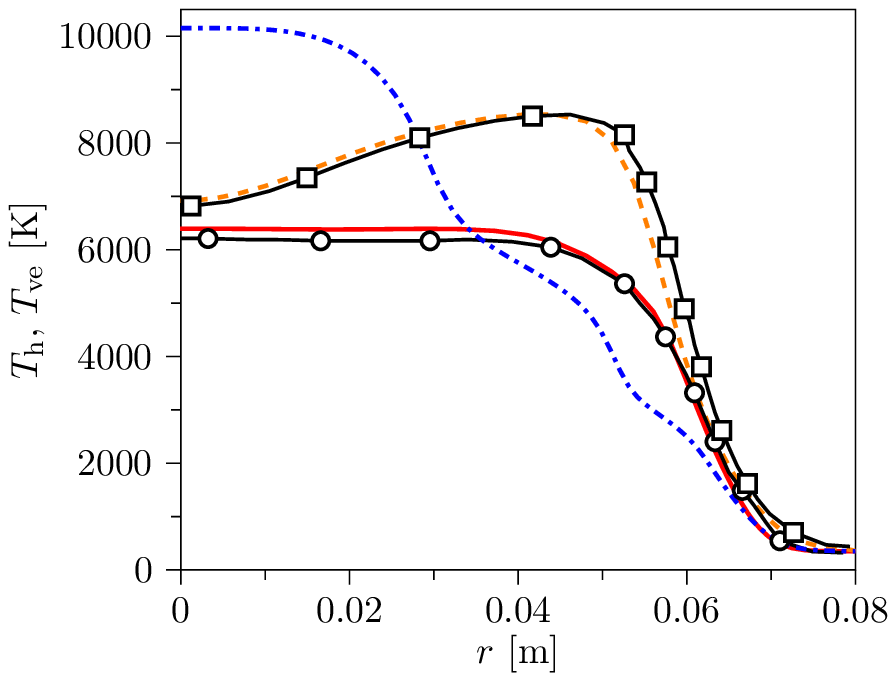} \\
	  \small (b) Temperatures ($x = \SI{0.235}{\meter}$).
  \end{tabular}
	\caption{ICP torch simulation (NLTE air plasma): in (a) temperature and normalized Joule heating distributions, in (b) temperature radial profiles at $x=\SI{0.235}{\meter}$. In (b) line heavy-particle temperature (this work), dashed line vibronic temperature (this work), line with circles heavy-particle temperature (Ref. \cite{David_thesis}), line with squares vibronic temperature (Ref. \cite{David_thesis}), dotted-dashed line LTE temperature (this work) ($\dot{m} = \SI{6}{\gram/\second}$, $f = \SI{0.45}{\mega\hertz}$, $P = \SI{50}{\kilo\watt}$, $p_{\mathrm{a}}= \SI{5000}{\pascal}$, $T_{\mathrm{w}} = \SI{350}{\kelvin}$; no swirl).}\label{fig:torch_nlte_ver}	
\end{figure}

\begin{figure}[!htb]
  \centering
  \begin{tabular}[b]{c}
    \includegraphics[clip,bb=0 3 250 195,keepaspectratio,height=0.355\textwidth]{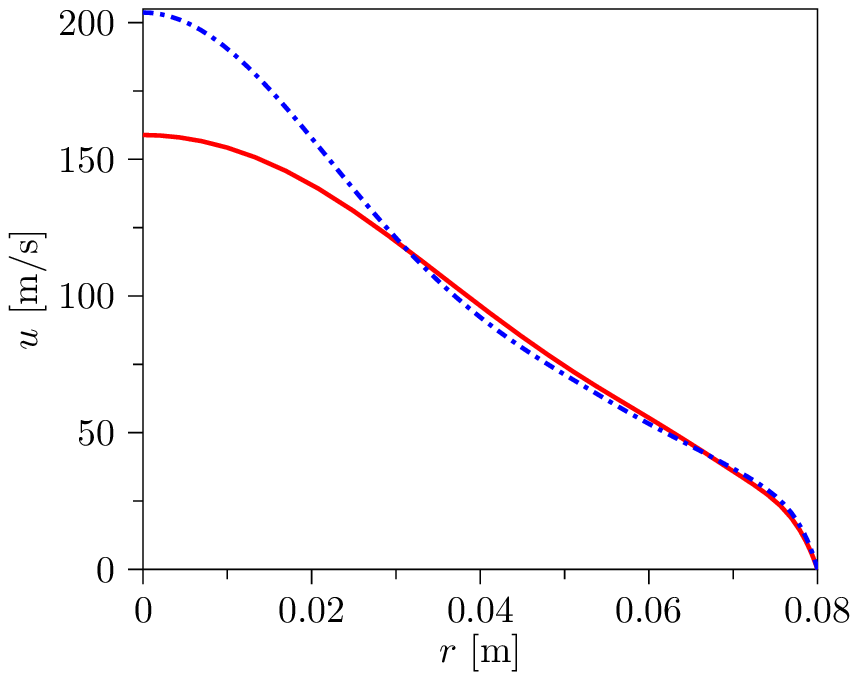} \\
    \small (a) Axial velocity.
  \end{tabular} \qquad
  \begin{tabular}[b]{c}
    \includegraphics[clip,bb=0 3 255 195,keepaspectratio,height=0.355\textwidth]{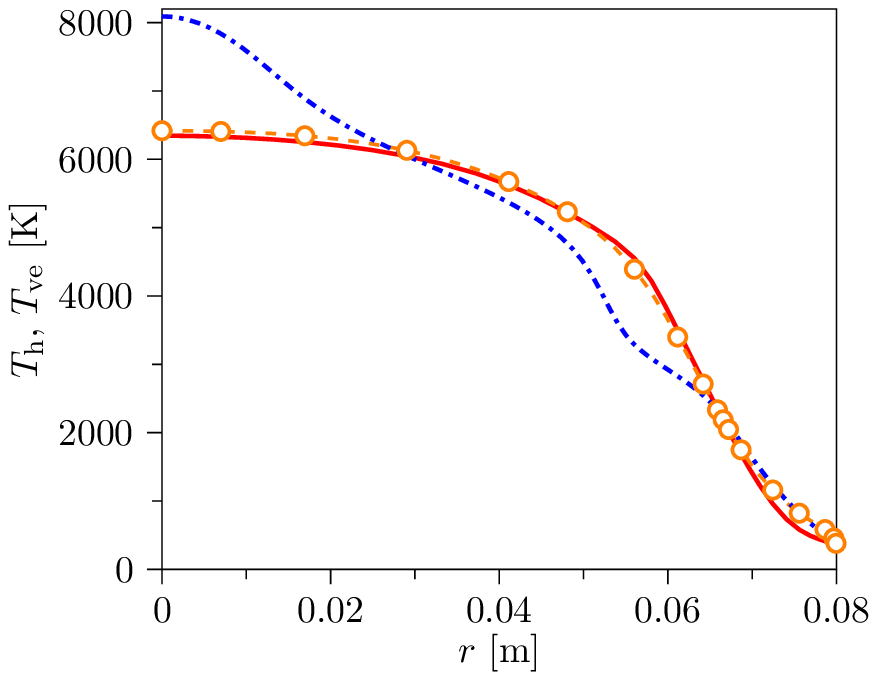} \\
	  \small (b) Temperatures.
  \end{tabular}
\caption{ICP torch simulation (NLTE air plasma): radial profiles of (a) axial velocity and (b) temperatures at torch exit ($x= \SI{0.47}{\meter}$). In (b) line heavy-particle temperature, dashed line with circles vibronic temperature. The dotted-dashed lines in both (a) and (b) denote the LTE simulation ($\dot{m} = \SI{6}{\gram/\second}$, $f = \SI{0.45}{\mega\hertz}$, $P = \SI{50}{\kilo\watt}$, $p_{\mathrm{a}}= \SI{5000}{\pascal}$, $T_{\mathrm{w}} = \SI{350}{\kelvin}$; no swirl).}\label{fig:torch_lte_nlte_exit}	
\end{figure}

Figure \ref{fig:torch_nlte_ver} shows the temperature distribution in the torch. Compared to the LTE simulation, the NLTE calculation leads to lower temperatures and a larger plasma volume \cite{Zhang_POP_ICP_2016}. Thermal non-equilibrium in the discharge is significant in the zone where the Joule heating is maximum (see the top of Fig. \ref{fig:torch_nlte_ver}(a)). Conversely, at the torch exit, the plasma is essentially in thermal equilibrium, though temperatures are significantly lower compared to the corresponding LTE values as shown in Fig. \ref{fig:torch_lte_nlte_exit} comparing LTE and NLTE axial velocity and temperature profiles. As for the LTE simulation, the results agree with the literature data. 

\section{Conclusions}
This paper has presented and discussed the development of a multi-physics framework for inductively coupled plasma (ICP) wind tunnels. As opposed to a monolithic approach, separate solvers responsible for the evolution of the plasma and the electromagnetic field have been coupled. The feasibility of the proposed methodology has been demonstrated for two-dimensional axisymmetric configurations. The implementation has been successfully verified via comparison against data available in the literature.  

Future work will focus on extending the framework to three-dimensional and unsteady scenarios, the inclusion of the radiation losses and TPS sample, as well as model validation through comparison against experiments performed at the University of Illinois.
\section*{ACKNOWLEDGEMENTS}
This work is supported by the Center for Hypersonics and Entry System Studies (CHESS) at the University of Illinois at Urbana-Champaign.
\bibliographystyle{aipnum-cp}
\bibliography{ref/ref,ref/ref2,ref/weno}
\end{document}